\documentclass[12pt]{iopart}
\def\beqra{\begin{eqnarray}}
\def\eeqra{\end{eqnarray}}
\def\beq{\begin{equation}}
\def\eeq{\end{equation}}

\begin{document}

\title{Non-linear gravitational clustering in scalar field cosmologies}

\author{Sabino Matarrese\dag\ddag, Massimo Pietroni\ddag\footnote[3]{To whom correspondence should be addressed}
and Carlo Schimd\P}
\address{\dag\ Dipartimento di Fisica `G. Galilei',
Universit\`{a} di Padova, \\ via Marzolo 8, I-35131 Padova.}
%\email{matarrese@pd.infn.it}
\address{\ddag\ INFN - Sezione di Padova, \\ via Marzolo 8, I-35131
Padova.}
%\email{pietroni@pd.infn.it}
\address{\P\ Dipartimento di Fisica `M. Melloni',
Universit\`{a} di Parma, \\ INFN -
Gruppo Collegato di Parma, \\ Parco Area delle Scienze 7/A, I-43100
Parma}
%\email{schimd@pr.infn.it}

\begin{abstract}
Non-linear gravitational clustering in a universe dominated by dark
energy, modelled by a `quintessence' scalar field, and cold dark matter
with space-time varying mass is studied. Models of this type, where
the variable mass is induced by dependence on the scalar
field, as suggested by string theory or extra-dimensions,
have been proposed as a viable solution of the coincidence problem.
A general framework for the study of the non-linear phases of
structure formation in scalar field cosmologies is provided, starting 
from a general relativistic treatment of the combined dark matter-dark energy 
system. As a first application, the mildly non-linear evolution of dark matter
perturbations is obtained by a straightforward extension of the Zel'dovich 
approximation. We argue that the dark energy fluctuations may play 
an active role in cosmological structure formation if the scalar field 
effective potential develops a temporary spinodal instability during the
evolution. 

\end{abstract}

\pacs{95.35.+d, 98.65+Dx,98.80.-k\\ DFPD 03/A/20, UPRF-2003-08}

\maketitle

\section{Introduction}

There is increasing agreement on the picture of the Universe on large
scales: the analysis of the magnitude-redshift relation for type Ia
Supernovae \cite{sn}, measurements of the
anisotropies of the Cosmic
Microwave Background (CMB) \cite{Spergel:2003cb},
and analyses of the galaxy distribution in large redshift catalogs
\cite{Verde:2001sf} all converge to show that the Universe is
spatially flat and composed by roughly 30\% of cold dark matter (DM),
responsible for the structure we see via the small amount of baryons,
and, for the remaining 70\% by an unclustered form of dark energy (DE),
responsible for its present-day accelerated expansion.
In order to alleviate or avoid fine-tuning of initial
conditions, an alternative solution to the usual cosmological constant
has been proposed in the form of a dynamical vacuum energy component with
negative equation of state, the so-called  quintessence
\cite{Caldwell:1998}.

The most common candidate for quintessence is a self-interacting
scalar field $\phi$. Different choices for the effective potential have 
been considered in the literature. In order to avoid fine tuning on the 
initial conditions, particular attention has been devoted to potentials 
leading to scaling or tracker behavior, like those depending on the 
scalar field exponentially \cite{Ferreira} or with a negative power-law 
\cite{Ratra}.

In addition to self-interactions, the DE scalar field might in principle 
be coupled to any other particle  or field present in Nature. However, 
given the extremely low mass typically expected for $\phi$ today, 
$m_\phi\sim H_0\sim 10^{-33} $ eV, its couplings to common particles, 
such as baryons or photons, are severely bounded by experimental tests 
on the equivalence principle and on the time variation of coupling 
constants.  On the other hand, the coupling to non-baryonic DM is not so 
tightly constrained and might play a significant role in the 
cosmological evolution \cite{Damour}. In this paper we will take into 
account such a possibility  by allowing the DM particles' mass to depend 
on the DE scalar field. The DM particles are then Varying Mass Particles 
(VAMPS), a possibility that was discussed in the past in connection to 
the age problem of the Universe \cite{Anderson:1997un}.

The VAMPS scenario was considered more recently  in the quintessence 
context in \cite{CPR}, where it was shown that it may lead to a solution 
of the so called `cosmic coincidence' problem. Indeed, assuming an 
exponential form both for the coupling and for the effective potential, 
the late time cosmology may be such that the universe accelerates and 
the energy densities of DE and DM scale at the same rate. As expected, 
in this scenario the predictions for the DM abundance are drastically 
altered with respect to the standard freeze-out scenario \cite{CPR}.
A coupling of this type may arise for radii moduli in braneworld 
scenarios \cite{Pietroni}. Couplings between DE and DM may also emerge 
in strongly coupled string theory \cite{GPV}.

It is generally believed that perturbations of the quintessence scalar
field play a negligible dynamical role in cosmic  structure formation,
because of the extreme smallness of the quintessence mass ($m_\phi\sim 
H$) which implies that its spatial
fluctuations only appear on very large scales and are bound to be linear.
Nevertheless, a number of models have been proposed where DE perturbations
might in principle grow non-linear and play an active role down to
galactic scales. These include coupled-quintessence \cite{Amendola:1999er},
extended quintessence \cite{Perrotta:1999am, Perrotta:2002sw},
inhomogeneous Chaplygin gas \cite{Bilic:2001cg}, k-essence
\cite{Armendariz-Picon:2000ah}.
A key question in the context of scalar field cosmologies is whether
fluctuations of the quintessence field can influence the galaxy formation
process in a sensible way, affecting the dynamics of structure formation
during the non-linear phases e.g. in the cores of dark matter halos.
This problem has been recently addressed by Wetterich 
\cite{Wetterich:2001},
but the issue of whether quintessence fluctuations can grow non-linear
on small scales and actively influence the non-linear clustering of
the CDM component is still completely open.

The VAMPS scenario considered here provides an useful bench mark to 
study situations where non-linearities in the scalar field may be 
induced either by the field's self interaction or by its couplings to 
other cosmic components.

The plan of the paper is as follows. In Section 2 we give a full
derivation of the equations which govern the non-linear evolution of DM 
and DE
perturbations on sub-horizon scales. Then, in order to gain some 
intuitive insight on the content of these equations, we specialize to 
the VAMP model discussed in \cite{CPR}. A simple solution of the
equations in the mildly non-linear regime for this model is obtained in 
Section 3, in
terms of first-order lagrangian perturbation theory. In Section 4 we discuss
how the simple model considered here could be modified to give rise to
non-negligible scalar field fluctuations on scales relevant for
galaxy or galaxy-cluster formation.

\section{General formalism}

As a starting point, we will expand the metric tensor $g_{\mu\nu}$
(greek indices run from 0 to 3, while latin ones from 1 to 3) to
linear order around a flat Robertson-Walker (RW) metric, while keeping
the full -- {\it i.e.} non-linear --
energy momentum tensor on the RHS of Einstein equations.
In the conformal Newtonian gauge,
  vector and tensor
perturbations of the metric decouple from the scalar ones as
long as they are treated linearly; by taking the spatial covariant
divergence of the $0-i$ Einstein equations and the trace of the $i-j$
equations, we get rid of vector and tensor modes and single out scalar
ones. Thus, it is not restrictive to consider the metric \cite{Ma:1995}
\begin{equation}
ds^2 = a^2(\eta)\left[ (1+2\Phi)d\eta^2 -(1-2\Psi)\delta_{ij}dx^i dx^j
\right] \, ,
\end{equation}
where the scalar perturbations $\Phi$ and $\Psi$ are related to the 
gravitational potential, and are
of order $|{\bf u}|^2\ll 1$ (${\bf u}$ being the typical velocity of
DM particles) even in the presence of highly non-linear DM overdensities.
Here $a(\eta)$ is the background scale-factor and $\eta$ is the
conformal time.

In this approximation, Einstein's equations read
\begin{equation} \label{eqn:hyb}
\delta R^\mu_\nu = 8\pi G (S^\mu_\nu - \bar{S}^\mu_\nu  )\, ,
\end{equation}
where $\delta R_{\mu\nu}$ is the linear perturbation  to the Ricci
tensor,
while $S^\mu_{\;\nu}
\equiv T^\mu_{\;\nu} - \frac{1}{2}\delta^\mu_{\;\nu} T$
is the fully non-linear source and $\bar{S}^\mu_\nu$ is the background
one.

The Friedmann equations for the background RW metric read
\begin{eqnarray}
\mathcal{H}^2 &=& \frac{8\pi G}{3}a^2 \left( \bar{\rho}_{m} +
\bar{\rho}_{\phi} \right) \label{fri1} \\
\dot{\mathcal{H}} &=& - \frac{4\pi G}{3}a^2 \left(\bar{ \rho}_{m} +
\bar{\rho}_\phi + 3\bar{p}_\phi \right)\label{fri2} \, ,
\end{eqnarray}
where $\mathcal{H}\equiv\dot{a}/a$, and the dot denotes
differentiation
with respect to $\eta$.

We will take into account the contribution to the energy-momentum
tensor coming from the dark matter and the scalar field. The former is made
up of a discrete set of particles with field-dependent mass
$m(\phi(\eta,{\bf x}))$ and coordinates ${\bf x}_a(\eta)$
($a=1,2,\ldots$), described by the energy--momentum tensor
\begin{equation}
T^\mu_{\;\nu}
\simeq  a^{-2}
\, \rho_m(\phi) \,u^\mu u_\nu \,,
\label{Tm}
\end{equation}
where
\[\rho_m(\phi)=m(\phi) \,a^{-3}\sum_{a}
\delta^{(3)}({\bf x} - {\bf x}_a)\,,
\]
$u^\mu\equiv dx^\mu/d\eta$ ($x^0\equiv\eta$),  and we have
neglected $O(|{\bf u}|^2)$ corrections.

The scalar field energy-momentum tensor reads
\begin{equation}
T^\mu_{\;\nu} = M_p^2 \phi^{;\mu}\phi_{\; ;\nu} -
\delta^\mu_{\;\nu}\left(
\frac{M_p^2}{2}\phi^{\; ;\alpha}\phi_{;\alpha} - V(\phi) \right) \;,
\label{Tphi}
\end{equation}
where $\phi$ is dimensionless because of the factorization of $M_p^2 =
(8\pi G)^{-1}$. After replacing the background quantities
$\bar{g}_{\mu\nu}$, $\bar{\phi}$, and $\bar{x}^\mu_a$ in (\ref{Tm})
and (\ref{Tphi}) one can read out the
background energy density and pressure for our two components,
$\bar{ \rho}_{m}$, $\bar{\rho}_\phi$, and $\bar{p}_\phi$.

The Einstein equations provide a redundant set of equations for the
evolution of the linearized metric perturbations; we choose to use the
$0-0$ equation and the spatial covariant derivative of the  $0-i$'s.
They read, respectively,
\begin{eqnarray}
\fl \nabla^2\Phi +3 \mathcal{H} (\dot{\Phi}+\dot{\Psi})+ 3 \ddot{\Psi}
=\frac{a^2}{2 M_p^2}
  \left[
\rho_m(\phi)
- \bar{\rho}_m
+ \frac{2 M_p^2}{a^{2}} (\dot{\phi}^2 -
\dot{\bar{\phi}}^2) - 2 ( V(\phi) - V(\bar{\phi}) ) \right]
\label{eqn:EC} , \\
\fl \mathcal{H}\nabla^2\Phi + \nabla^2\dot{\Psi} = \frac{a^2}{2 M_p^2}
\left[
\partial_i \left(\rho_m(\phi)\, u^i \right)+
\frac{M_p^2}{a^{2}}
(\partial_i\dot{\phi}\,\partial^i\phi + \dot{\phi}\,
\nabla^2\phi)+\frac{2 M_p^2}{a^{2}}
\dot{\phi}\,\partial_i\Psi\,\partial^i\phi \right] , \label{eqn:MC}
\end{eqnarray}
where primes denote differentiation with respect to $\phi$ and, from
now on, latin indices are raised with a Kronecker delta.

The evolution of the scalar field $\phi(\eta,{\bf x})$ is given by the
Klein--Gordon equation, which, for $|\Phi|, |\Psi|, |u^i| \ll 1$, reads
\begin{equation}\label{eqn:KG}
\ddot{\phi} + 2\mathcal{H}\dot{\phi} - \nabla^2\phi +
\frac{a^2}{M_p^{2}} U^\prime(\phi)
= \nabla(\Phi - \Psi) \cdot \nabla\phi \,,
\end{equation}
where the effective potential $U(\phi)$ is the sum of the scalar field
potential and the field-dependent dark matter energy density,
\begin{equation}
U(\phi) \equiv V(\phi) + \rho_{m}(\phi)
\end{equation}

Finally, we need the equation of motion for the dark matter particles,
\begin{equation*}
\frac{d^2 x^\mu}{d s^2}+ \Gamma^\mu_{\sigma\beta} \frac{d
x^\sigma}{ds}
\frac{d x^\beta}{ds} - \frac{m^\prime}{m} \partial^\mu\phi = 0 \,,
\end{equation*}
where the last term comes from the field dependence of the particle
mass and makes the trajectories deviate from geodesics.
The spatial coordinates of the $a$-th particle
satisfy \begin{equation}\label{eqn:EoM}
\ddot{x}_a^i + \dot{x}_a^i \left[ \, \mathcal{H} +
\frac{m^\prime}{m}\dot{\phi} \, \right] = - \partial^i \Phi -
\frac{m^\prime}{m} \partial^i\phi \,,
\end{equation}
where $\mathcal{H}$, $\Phi$, and $\phi$ are evaluated at
$(\eta,{\bf x}_a(\eta))$.

Equations (\ref{eqn:EC}), (\ref{eqn:MC}), (\ref{eqn:KG}) and
(\ref{eqn:EoM}) form a close system in the variables $\Phi$, $\Psi$,
$\phi$ and $x_a^i$'s.

As we are interested in sub-horizon scales, there are further
approximations which can be made. First of all, in Eq.~(\ref{eqn:EC}),
only the first term on the LHS will be kept, as the others are
${\cal O}(\mathcal{H}^2 \Phi\,,\mathcal{H}^2 \Psi)$. Furthermore, the RHS
of Eq.~(\ref{eqn:KG}) is suppressed by at least an
${\cal O}(\Phi,\Psi)$ factor with respect to the Laplacian on the LHS, and
can then
be dropped. As a result, the $\Psi$ perturbation is
not needed anymore, $\Phi$ can be identified with the gravitational
potential, and we can concentrate on the reduced system of
Eqs.~(\ref{eqn:EC}), (\ref{eqn:KG}), and (\ref{eqn:EoM}) in their
approximated forms.

To illustrate the role of the scalar field we split it  into
background, `non-relativistic'  and `relativistic' contributions,
\begin{equation}
\phi(\eta,{\bf x}) = \bar{\phi}(\eta) + \phi_{NR}(\eta,{\bf x}) +
\phi_{R}(\eta,{\bf x}) \;.
\end{equation}
$\bar{\phi}(\eta)$ is the solution of the background Klein-Gordon
equation,
\begin{equation}
\ddot{\bar{\phi}} + 2\mathcal{H}\dot{\bar{\phi}} +
a^2M_p^{-2}U^\prime(\bar{\phi}) = 0 \label{KGback} \,,
\end{equation}
while the non-relativistic part $\phi_{NR}(\eta,{\bf x})$ is defined
as the solution of the Poisson-like equation
\begin{equation}
\nabla^2\phi_{NR} = a^2M_p^{-2}\Delta U^\prime(\bar{\phi} +
\phi_{NR})\, ,
\end{equation}
where $\Delta U(\phi) \equiv U(\phi) - V(\bar{\phi}) - \bar{\rho}_m$.
Thus, $\phi_{NR}$ accounts for the response of the scalar field to the
localized DM distribution, as induced by the field dependence of the DM
particle mass. It behaves similarly to the gravitational potential $\Phi$
and, as we will see, in specific cases it may turn out to be just
proportional to the latter.

The remaining `relativistic' part solves the equation
\begin{equation}
\fl\ddot{\phi}_R + 2\mathcal{H}\dot{\phi}_R - \nabla^2\phi_R +
\ddot{\phi}_{NR} + 2\mathcal{H}\dot{\phi}_{NR}  +
a^2M_p^{-2}
[\Delta U^\prime(\bar{\phi} + \phi_{NR} + \phi_{R}) - \Delta
U^\prime(\bar{\phi} + \phi_{NR})]=0\,.
\label{phirel}
\end{equation}

The effective mass for $\phi_R$, as read out from Eq.~(\ref{phirel})
is given by $\Delta U^{\prime\prime}(\bar{\phi}+\phi_{NR})$,
which, for the scales of interest for the formation
of cosmic structures is much smaller than the
corresponding $k^2/a^2$. Moreover, since the time-scale for
$\bar{\phi}$ and $\phi_{NR}$ variation is given by $\mathcal{H}$, the
$\dot{\phi}_{NR}$ and $\ddot{\phi}_{NR}$ terms in (\ref{phirel})
provide ${\cal O}(\mathcal{H}^2)$ sources.
As a consequence, the $\phi_R$ component behaves essentially as the
superposition of approximately massless plane waves on the scales of
interest, with ${\cal O}(a^2\bar{\rho}_m/k^2 M_p^2)\ll1$ amplitudes, and can
then be neglected.

The relevant dynamics can then be described in terms of the equations

\begin{eqnarray}
&&\nabla^2\Phi = \frac{a^2}{2 M_p^2}
\,\left[\rho_m(\bar{\phi}+\phi_{NR})- \bar{\rho}_m
-2
\left(V(\bar{\phi}+\phi_{NR})-V(\bar{\phi})\right)\right]\,,
\label{POISSON} \\
&&\nabla^2\phi_{NR} = \frac{a^2}{M_p^{2}}\Delta U^\prime(\bar{\phi} +
\phi_{NR}) \,,
\label{KGNR}\\
&&
\ddot{x}_a^i +  \left[ \mathcal{H}
+\frac{m^{\prime}}{m}(\dot{\bar{\phi}}+\dot{\phi}_{NR})\right]
\,\dot{x}_a^i = - \partial^i \Phi - \frac{m^{\prime}}{m} \partial^i
(\bar{\phi} + \phi_{NR})
\,, \label{eqn:EoMS}
\end{eqnarray}

and the background ones, Eqs.~(\ref{fri1}), (\ref{fri2}), and
(\ref{KGback}).

The equations above extend the standard Newtonian approximation from the 
pure DM case to the coupled system of DM and quintessence DE. All 
non-background DE effects are given by $\phi_{NR}$, which solves the 
Poisson-like eq.~(\ref{KGNR}).

In order to get a flavour of the dynamical content of the system, 
we specialize to a simple model which was proposed as a
solution of the coincidence problem \cite{CPR, Pietroni}.
The scalar potential and the DM particle mass are
both exponentially dependent on $\phi$,
\begin{equation}
V(\phi)= \hat{V} \exp(\beta \phi)\,,\,\,\,\,m=\hat{m} \exp(-\lambda
\phi)\,,\label{expo}
\end{equation}
with $\beta , \,\lambda>0$.
The model has a solution such that
\begin{equation}
\bar{\phi} = \frac{-3}{\lambda+\beta}\;\log
a\,,\;\;\;\;\;\;\;\;
\Omega_\phi = 1-\Omega_m =
\frac{3+\lambda(\lambda+\beta)}{(\lambda+\beta)^2}\, ,
\label{attractor}
\end{equation}
which is an {\em attractor} in field space if
$\beta>(-\lambda+\sqrt{\lambda^2+12})/2$. From the moment
the attractor is reached, the energy densities in DM
and in DE evolve at a constant ratio depending only on $\lambda$ and
$\beta$, thus solving  the cosmic coincidence problem.

Indeed, the  solution (\ref{attractor}) implies
\begin{equation}
\bar{\rho}_m = m(\bar{\phi}) \bar{n}_m \sim \bar{\rho}_\phi \sim
a^{-3(1+w)}\,,
\;\;\mbox{with}\;\;\;w=-\frac{\lambda}{\beta+\lambda}\,,
\end{equation}
where $\bar{n}_m$ is the DM background number density. The negative
$w$ leads, if $w<-1/3$, to accelerated expansion of the universe.
The $\phi$ dependence of the DM mass modifies the usual scaling
$a^{-3}$ of non relativistic matter. Since the mass increases
with the expansion, the effect is analogous to that of a fluid with
negative equation of state, though  DM is, as in the standard
scenario,  a pressureless fluid made up by non-interacting particles.

The impact on the CMB of this model is similar to the one considered
in \cite{Amendola:2000uh}. It was shown there that in order to get 
results in
agreement with the CMB anisotropy spectrum a time-dependent $\lambda$
should be invoked, such that the early time dependence is the standard
CDM one, and the attractor behavior of Eq.~(\ref{attractor}) starts
only recently. Since we use the model only for illustrative purposes,
we will stick to a constant $\lambda$ in the following.

The exponential dependences on $\phi$ in $\Delta U$ and
Eq.~(\ref{KGNR}) imply that also $\phi_{NR}={\cal O}(a^2\bar{\rho}_m/k^2
M_p^2)\ll1$. Then, as dark matter fluctuations approach the non linear
regime, they start dominating over those of the effective potential,
{\it i.e.}
\begin{equation}
\Delta U \sim m(\bar{\phi})\left(a^{-3}  \sum_a \delta^{(3)}({\bf x }-
{\bf x}_a)-\bar{n}_m\right)\,.
\end{equation}
In this approximation, the RHS of the Poisson's equation
(\ref{POISSON}) and that of Eq.~ (\ref{KGNR}) are proportional to each
other, so that $\phi_{NR} = -2\lambda \Phi=O(|\bf{u}|^2)$.

Using the properties of the attractor solution, Eq.~(\ref{attractor}),
we get the further reduced system,
\begin{eqnarray}
\nabla^2\Phi& =& \frac{\hat{m} }{2 M_p^2}
\,a^{2-3w}\left(a^{-3}\sum_{a} \delta^{(3)}({\bf x} - {\bf
x}_a)-\bar{n}_m\right)\,, \label{eqn:PoissonGrav} \\
\ddot{x}_a^i &+&  (1-3 w) \mathcal{H} \,\dot{x}_a^i
= - (1+2\lambda^2)\,
\partial^i \Phi \label{geomod}
\end{eqnarray}
The coupling between DM and DE modifies the equations from the
standard ones in two ways. It  changes the background evolution
through the $w$-terms, and it gives an extra contribution to the force
between two DM particles, due to the exchange of a light $\phi$
quantum. The last effect is responsible for the $2 \lambda^2$
correction to the gravitational force in the LHS of
Eq.~(\ref{geomod}).

\section{Mildly non-linear evolution}

The set of equations (\ref{eqn:PoissonGrav}), (\ref{geomod})
provides the required extension of the standard Newtonian approximation
(e.g. \cite{Peebles:1980}), describing the non-linear
evolution of matter perturbations, to the case where a quintessence
scalar field component is present. As a preliminary application, we can 
study
the effect of the scalar field on the DM evolution in the midly non-linear
regime, using first-order Lagrangian perturbation theory.

Standing to standard notation \cite{zel}, one can look
for a time variable $\tau = \tau(\eta)$ by which the equation of motion
derived from (\ref{eqn:EoMS}) does not contain the Hubble drag term.
We can then write an equation for the displacement from the initial
(Lagrangian) position $\mathbf{q}$ to the final (Eulerian) position
$\mathbf{x} = \mathbf{q} + \mathbf{S}(\tau,\mathbf{q})$ subjected to
a force per unit mass proportional to $\partial^i \Phi$.
After taking the divergence of the resulting equation of motion, one
can use the Poisson Eq. (\ref{eqn:PoissonGrav})  to
obtain an equation for the {\it deformation tensor}
$\mathcal{S}_{ij}(\tau,\mathbf{q}) \equiv
\partial S^i/\partial q^j$ and then reconstruct the density
field $\rho(\tau , \mathbf{x}) = \rho(\mathbf{q})J^{-1}$, where
$J=\det(\delta_{ij} + \mathcal{S}_{ij})$ is the determinant of the
Jacobian $\partial x^i/\partial q^j$.

On the attractor (\ref{attractor}),
for $\tau = a^n$ and $n=-\frac{1}{2}(1+9w)$, the DM
equation of motion is simplified to
\begin{equation} \label{eqn:EoMcv}
\frac{d^2 \mathbf{S}}{d\tau^2} = -
\frac{(1+2\lambda^2)}{n^2\tau^2\mathcal{H}^2}\nabla_\mathbf{x} \Phi
\equiv
- \nabla_\mathbf{x} \varphi
\end{equation}
and the Poisson equation (\ref{eqn:PoissonGrav}) becomes
\begin{equation}
\label{eqn:LPT_first} \nabla^2_\mathbf{x}\varphi = \left(
\frac{3}{2}\Omega_m \frac{(1+2\lambda^2)}{n^2\tau^2} \right) \delta_m
(\mathbf{x}(\tau,\mathbf{q})) \equiv b(\tau) \delta \, ,
\end{equation}
where $\delta = \delta_m(\mathbf{x}(\tau,\mathbf{q}),\tau) =
[\rho_m (\tau,\mathbf{x}) - \bar{\rho}_m (\tau)]/ \bar{\rho}_m (\tau)$
is the density contrast. 

Expanding the resulting equation to first order in the
displacement vector $\mathbf{S}$ and making the ansatz
$\mathbf{S}(\tau,\mathbf{q}) = f(\tau)\mathbf{\tilde{S}}(\mathbf{q})$
we obtain
\begin{equation}
\frac{d^2}{d\tau^2} f(\tau) - b(\tau) f(\tau) = 0 \, ,
\end{equation}
whose solutions read
\begin{equation}
f(\tau) = \tau^p \, , \quad p = \frac{1}{2} \pm \frac{1}{2}\sqrt{1+\frac{24
\Omega_m(1+2\lambda^2)}{(1-9w)^2}}\, .
\end{equation}
We also get $\nabla_{\bf q} \cdot \mathbf{\tilde{S}}(\mathbf{q}) = -
\delta_0$, where $\delta_0$ is the VAMPS density fluctuation linearly extrapolated to the present time.

The matter density field follows:
\begin{equation}
\rho_m(z,\mathbf{x}) = \frac{\bar{\rho}_m (1+z)^{3(1+w)}}
{\prod_{i=1}^{3} \left[ 1 +
\lambda_i(\mathbf{q})(1+z)^{-m_+} \right] } \, ,
\end{equation}
where
$\lambda_i(\mathbf{q})$ are the eigenvalues of $\partial \tilde{S}_j/
\partial q^i$ and
\begin{equation}
m_\pm = -\frac{1}{4}(1-9w)\left(
\mp \sqrt{1+\frac{24 \Omega_m (1+2\lambda^2)}{(1-9w)^2}} \right)
\end{equation}
(in agreement with \cite{Amendola:2000uh}): the effect of the
scalar field and a non-vanishing coupling constant $\lambda$ is to 
anticipate
(if $m_+ > 1$) or delay (if $m_+ < 1$) the formation of pancakes with
respect to scenarios with only DM with constant mass ($m_+ = 1$),
depending on the value of $\lambda$ and $\beta$.
Because the DM-DE coupling in our specific model affects 
the motion of DM particles only through a $2\lambda^2$ correction 
of the gravitational force, the eingenvectors of the 
deformation tensor $\partial \tilde{S}_j/ \partial q^i$ turn out to be exactly 
aligned with those of the gravitational force itself.
As a consequence, no other scalar field effects appear on small 
scales.   

\section{Conclusions}

In this paper we have set the stage for future studies of non-linear 
gravitational clustering in scalar field cosmologies, extending the 
Newtonian approximation to the case of a coupled DM-DE system. The 
non-linearity in the scalar field may emerge either as a result of its 
self interactions or by its coupling to DM. It obeys a Poisson-like 
equation, Eq.~(\ref{KGNR}), quite similar to that for the gravitational 
potential $\Phi$.

In order to get some analytic insight, we have specialized to the VAMP 
scenario discussed in \cite{CPR}, in which the DM particle's mass depends exponentially on the  DE field. In this particular model, we found that the scalar field perturbations remain linear inside the horizon, although the coupling plays a non-trivial dynamical role.

The mildly non-linear regime has been tested here, using first-order Lagrangian perturbation theory. The DM-DE coupling weakly affects the gravitational field, acting on the onset of pancake formation without any further consequence on the spatial distribution of the DM density field on the scales of interest.

The constant coupling $\lambda$ considered here is, however, disfavored by comparison with the CMB anisotropy spectrum. A time- or field-dependent $\lambda$ -- smaller in the past and approximately constant in the recent epoch -- would take care of this problem, as discussed in \cite{Amendola:2000uh}. In this case, our results would not change qualitatively. Namely, the scalar field fluctuations would still remain well inside the linear regime and the onset of structure formation would be changed with respect to ordinary quintessence, though of a lesser extent than for a constant $\lambda$.

Scalar field non-linearities are, however, not precluded in the VAMP scenario. For instance, modifying the field dependence of the DM mass in Eq.~(\ref{expo}), the full potential $U(\phi)$ may temporarily develop a spinodal instability (negative second derivative). During this time, perturbations at scales $k/a \le \sqrt{-U^{\prime\prime}}$ grow exponentially and easily exit the linear regime. When the spinodal epoch ends, the system can be described by our equations (\ref{POISSON}), (\ref{KGNR}), (\ref{eqn:EoMS}), assigning the developed non-linearities to the initial conditions for the non-relativistic field fluctuation $\phi_{NR}$.

The approach developed in this paper can be straightforwardly applied to 
the numerical study of models where the quintessence field fluctuates 
non-linearly, as those proposed in refs.  \cite{Amendola:1999er, 
Perrotta:1999am, Perrotta:2002sw, Bilic:2001cg, Armendariz-Picon:2000ah, Wetterich:2001}.

\subsection{Acknowledgments}
We thank F. Perrotta for useful discussions.

\end{document}